\documentclass[journal=jacsat,manuscript=article]{achemso}

\usepackage{chemformula} 
\usepackage[T1]{fontenc} 
\usepackage[figuresright]{rotating}

\author{Shuke Zhang}
\affiliation{Software College, Hebei Normal University, Shijiazhuang 050024, China}
\alsoaffiliation{Shijiazhuang Xianyu Digital Biotechnology Co., Ltd, Shijiazhuang 050024, China}
\altaffiliation{S.Z. and Y.J. equivalent authors}
\author{Yanzhao Jin}
\affiliation{Software College, Hebei Normal University, Shijiazhuang 050024, China}
\alsoaffiliation{Shijiazhuang Xianyu Digital Biotechnology Co., Ltd, Shijiazhuang 050024, China}
\altaffiliation{S.Z. and Y.J. equivalent authors}
\author{Tianmeng Liu}
\affiliation{Software College, Hebei Normal University, Shijiazhuang 050024, China}
\alsoaffiliation{Shijiazhuang Xianyu Digital Biotechnology Co., Ltd, Shijiazhuang 050024, China}
\author{Qi Wang}
\affiliation{Software College, Hebei Normal University, Shijiazhuang 050024, China}
\alsoaffiliation{Shijiazhuang Xianyu Digital Biotechnology Co., Ltd, Shijiazhuang 050024, China}

\author{Zhaohui Zhang}
\email{zhangzhaohui@hebtu.edu.cn}
\affiliation{College of Computer and Cyber Security, Hebei Normal University, Shijiazhuang 050024, China}
\alsoaffiliation{Software College, Hebei Normal University, Shijiazhuang 050024, China}

\author{Shuliang Zhao}
\email{shuliangzhao@hebtu.edu.cn}
\affiliation{College of Computer and Cyber Security, Hebei Normal University, Shijiazhuang 050024, China}
\alsoaffiliation{Hebei Provincial Key Laboratory of Network and Information Security, Shijiazhuang 050024, China}
\alsoaffiliation{Hebei Provincial Engineering Research Center for Supply Chain Big Data Analytics $\&$ Data Security, Shijiazhuang 050024, China}

\author{Bo Shan}
\email{shanbo@onest.net}
\affiliation{Software College, Hebei Normal University, Shijiazhuang 050024, China}
\alsoaffiliation{Shijiazhuang Xianyu Digital Biotechnology Co., Ltd, Shijiazhuang 050024, China}

\title[An \textsf{achemso} demo]
  {SS-GNN: A Simple-Structured Graph Neural Network for Affinity Prediction}

\abbreviations{IR,NMR,UV}
\keywords{American Chemical Society, \LaTeX}

\begin{document}
\captionsetup[figure]{labelfont={bf}}






\begin{abstract}

Efficient and effective drug-target binding affinity (DTBA) prediction is a challenging task due to the limited computational resources in practical applications and is a crucial basis for drug screening. Inspired by the good representation ability of graph neural networks (GNNs), we propose a simple-structured GNN model named SS-GNN to accurately predict DTBA. By constructing a single undirected graph based on a distance threshold to represent protein–ligand interactions, the scale of the graph data is greatly reduced. Moreover, ignoring covalent bonds in the protein further reduces
the computational cost of the model. The GNN-MLP module takes the latent feature extraction of atoms and edges in the graph as two mutually independent processes. We also develop an edge-based atom-pair feature aggregation method to represent complex interactions and a graph pooling-based method to predict the binding affinity of the complex. We achieve state-of-the-art prediction performance using a simple model (with only 0.6M parameters) without introducing complicated geometric feature descriptions.
SS-GNN achieves Pearson’s $R_{p}$=0.853 on the PDBbind v2016 core set, outperforming state-of-the-art GNN-based methods by 5.2$\%$. Moreover, the simplified model structure and concise data processing procedure improve the prediction efficiency of the model. For a typical protein–ligand complex, affinity prediction takes only 0.2 ms. All codes are freely accessible at https://github.com/xianyuco/SS-GNN.
\end{abstract}

\section{1 Introduction}
Drug development is a process with long cycles, high investments and high risks.\cite{mullard2014new,ashburn2004drug}. 
Drug-target binding affinity (DTBA) prediction plays an important role in drug development\cite{thafar2019comparison,chen2016drug,mei2019multi,alshahrani2018drug} and is also an important basis for drug screening.
Accurate DTBA predictions will significantly reduce new drug development costs and speed up the drug discovery process \cite{ozturk2019widedta}, which remains a challenge today.
Traditional methods such as classical scoring functions (SFs)\cite{guedes2018empirical,huang2006iterative,liu2013knowledge,li2019overview} do not estimate binding affinity well, and molecular dynamics (MD) simulations \cite{king2021recent,abel2017advancing} have improved prediction accuracy, but they are too slow for large-scale applications.
With the development of machine learning (ML), a large number of models for predicting drug-target interactions based on traditional ML methods\cite{yamanishi2010drug,nascimento2016multiple,cheng2016effectively,ballester2014does,li2015improving, wang2017improving,durrant2010nnscore} have emerged.
$\Delta \mathrm{VinaRF}_{20}$ combines AutoDock Vina and random forest models to predict binding affinity, and AGL-Score\cite{nguyen2019agl} and HPC/HWPC\cite{liu2021hypergraph} are based on algebraic graph descriptors and topological descriptors for molecular representation, respectively, and are trained via gradient boosted trees (GBT). These ML models have achieved good results; however, they typically use well-designed manual features and require special domain knowledge and experience. In addition, their learning ability and generalization have certain limitations.

Deep learning (DL)-based methods can automatically extract features from available data. Therefore, DL-based methods have received increasing attention, and a large number of DL-based methods\cite{karimi2019deepaffinity,ozturk2018deepdta,rogers2010extended,liu2019chemi,ozturk2019widedta}  have been proposed for binding affinity prediction, most of which have better performance and greater potential for capacity enhancement than traditional ML algorithms. Among them, the most commonly used methods are convolutional neural networks (CNNs) and graph neural networks (GNNs). With the increase in ligand-target 3D structural data, learning to predict binding affinity from 3D structural complexes has become a hot area of research. To encode the structural information of proteins and drugs as comprehensively as possible, some DL models based on 3D structure embedding\cite{lim2019predicting,hassan2020rosenet,ragoza2017protein,stepniewska2018development,wallach2015atomnet} have been proposed.
In OnionNet\cite{zheng2019onionnet}, the contacts between proteins and ligands are grouped according to different distance ranges, and the resulting features are fed into a CNN. In $\mathrm{K_{Deep}}$\cite{jimenez2018k}, FAST\cite{jones2021improved}, and Pafnucy\cite{stepniewska2018development}, 3D grids are applied to represent protein–ligand complexes, and 3D CNNs are applied to generate feature embeddings.

Although CNN-based methods have achieved remarkable progress in DTBA prediction, most models may have difficulty representing molecular graph structure features well. To better solve this problem, GNNs with good representation ability are used for DTBA prediction.\cite{wu2020comprehensive,nguyen2021graphdta, niepert2016learning,gao2018large,sun2020graph,lim2019predicting,zhou2020distance,klicpera2020directional}. 
In GNN-based models, graph structures are applied to represent atoms and their covalent bonds, and GNNs are applied to predict drug-target binding affinity. Some new methods have been proposed that consider the spatial information of the relative positions of atoms between ligands and proteins to improve GNN-based DTBA prediction models. The SGCN model\cite{danel2020spatial} proposed by Danel et al. considers the spatial information of the nodes. SIGN\cite{li2021structure} proposed by Li et al. introduces polar coordinates and considers angle information and the case of long-range interactions between atoms.
The IGN model proposed by Jiang et al.\cite{jiang2021interactiongraphnet} 
encodes the chemical and structural information in 3D space into a molecular graph, which comprehensively represents protein–ligand interaction patterns and adopts three molecular graphs to represent complexes. The model has good performance on the PDBbind dataset. GNN-based frameworks considering 3D structural information have made good progress in binding affinity prediction, but most of these frameworks employ intricate geometric structures that complicate the model. They are still not well suited for downstream molecular docking techniques for practical large-scale virtual screening (VS).
Therefore, it is highly desirable to develop an efficient DTBA prediction model with simple geometric structure information to meet the requirement of high efficiency.

To tackle the above problems, in this paper, we develop a novel method to improve the DTBA prediction model based on a GNN named SS-GNN. Compared with the state-of-the-art methods, it not only achieves good prediction performance but also has a simple structure and high prediction efficiency. SS-GNN is equipped with three modules to accomplish affinity prediction. We apply a single undirected graph to represent protein–ligand complexes, where nodes are atoms and edges are the interactions of atoms (Figure 1).
\begin{figure}
   \centering
   \includegraphics[width=12cm]{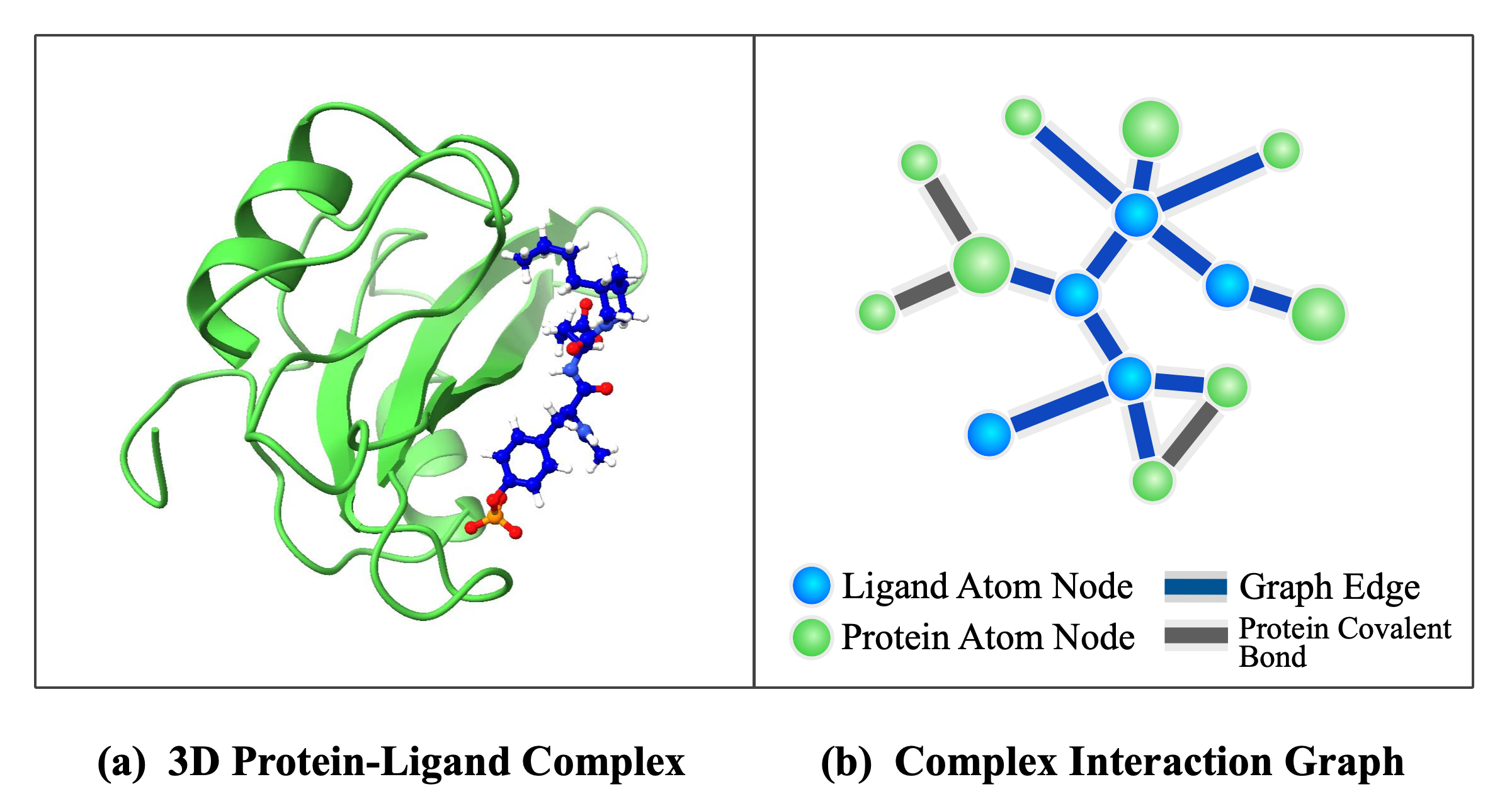}
  \caption{\textbf{Graph representation of the protein–ligand complex. (a) 3D structure of the complex. (b) Graph representation ignoring protein atoms outside the threshold and all covalent bonds in the protein.}}
\end{figure}
The graph representation method, which introduces a distance threshold, greatly reduces the size of the data. We design a hybrid feature extraction module (GNN-MLP) to extract useful features for atoms and interactions, respectively, and implement a lightweight feature embedding process via a 2-layer GIN submodule and a 3-layer MLP submodule. By aggregating the embedding information of each edge and its connected atom pairs, edge-based atom-pair aggregation features can be obtained, and by applying a simple MLP, the binding affinity of a single edge can be predicted. Finally, by summing the individual edge affinity predictions by employing a graph pooling module, the affinity of the complex can be obtained. In summary, the main contributions of our work are as follows:

\textbf{(1)	Protein–ligand complex representation based on a single undirected graph.} 
By means of a cross-validation-based distance threshold selection strategy, the protein atoms that are far away from the ligand can be removed, and some unimportant structures can be effectively avoided. Ignoring the covalent bonds in the protein further reduces the data size. The model achieves the best trade-off between prediction accuracy and computational complexity. The discretization of the interatomic distance improves the computational efficiency and generalization ability to a certain extent and further improves the performance of the model.

\textbf{(2)Hybrid feature extraction based on GNN-MLP.} We regard the feature extraction of atoms and edges in the graph as two independent processes: the atom features are extracted by applying a simple and effective 2-layer GIN, and the edge features are extracted by applying a lightweight MLP. Moreover, the single undirected graph representation not only simplifies the model but also makes updating the node information of proteins and ligands in the GNN more efficient.

\textbf{(3)	Edge-based atom-pair feature aggregation and graph pooling-based affinity prediction.} The embedding vectors of each edge and its connected atom pairs are concatenated to achieve feature aggregation and form the inputs of the affinity prediction module. The prediction outputs of all individual edges are summed through a graph pooling layer to obtain the binding affinity of the complex.

Unlike other models, our data processing procedure avoids the high complexity caused by extracting complicated geometric structures, and the absence of edges between protein atoms drastically reduces the computation. As a result, the number of parameters in the entire model is only 0.6M. The simplicity of the model and data processing procedure lead to a simple and low-complexity SS-GNN. Experiments demonstrate the effectiveness and efficiency of the proposed model. In Section 2, we introduce the detailed model architecture of SS-GNN. In Section 3, we present the experimental results and compare them with those of state-of-the-art methods in similar tasks. Finally, Section 4 summarizes our proposed method and briefly describes our future research plans.

\section{2 SS-GNN}
In this section, we introduce the proposed SS-GNN method. The SS-GNN defines the prediction of DTBA as a regression task, in which the model’s input is the drug-target representation, and the output is a continuous value representing the binding affinity score between the drug and the target protein. The overall architecture of the SS-GNN is shown in Figure 2. Our approach consists of graph representation of complexes based on the distance threshold, hybrid mode feature extraction, feature aggregation and affinity prediction. We first give an overview of the SS-GNN considering the 3D structure of the protein–ligand complex. In the following subsections, we elaborate the key modules.
\begin{sidewaysfigure}
   \centering
   \includegraphics[width=22cm]{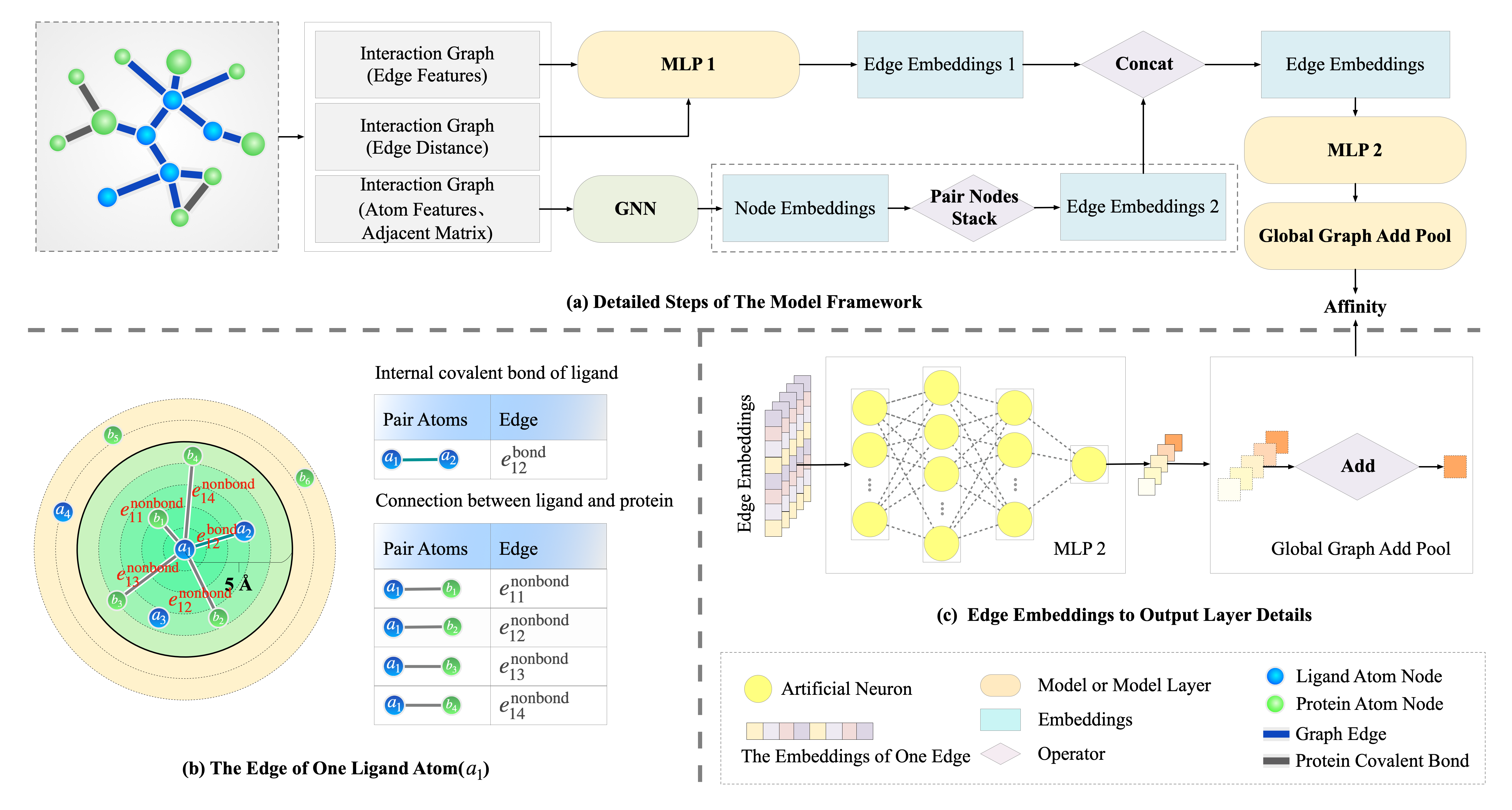}
  \caption{\textbf{(a) Detailed steps of the SS-GNN framework. The SS-GNN takes a graph representation of drug-protein complexes as input and the prediction of binding affinity as output. (b) The two types of edges connected to an example ligand atom. (c) Details of the affinity prediction module.}}
\end{sidewaysfigure}

\subsection{2.1 Protein–ligand Complex Representation based on a Single Undirected Graph}
Given a protein–ligand complex as shown in Figure 1(a), it can be described by the graph of interactions between atoms. As a rule of thumb, when the distance between protein–ligand atom pairs is greater than a certain threshold, the interactions between them do not contribute much to the interactions of the overall complex. In the initial stage of the experiment, we have constructed the complex graph using ligand atoms as well as all protein atoms and achieved good results. However, the number of atoms in a protein is much larger than that in a ligand. 
To verify whether the ligand features are overwhelmed by excessive protein features, we remove all the features of the ligand and replace them with random numbers. Experiments show that the model is still valid to a certain extent, which has led us to think about how to make better use of ligand features and whether all protein atoms are required in the model. To this end, we propose a distance-threshold-based graph representation method that employs the ligand and its partial protein to construct a complex interaction graph. We set the distance threshold as an optimizable hyperparameter and experimentally determine a feasible value.

Unlike some other GNN-based DTBA prediction methods, our proposed SS-GNN applies only one single protein–ligand complex graph $\mathcal{G}$ to characterize the interactions of the complex instead of building ligand and protein graphs separately. 
We first define the atom node set of the ligand as $\mathcal{V}^{L}$. We also define the atom node set of the protein as $\mathcal{V}^{P}=\left\{ a_{j} |\ a_{j} \ \textrm{is\ a\ protein\ atom\ satisfying}\  d\left ( a_{i},a_{j} \right ) \leq \theta,\ \textrm{where}\ a_{i}\in \mathcal{V}^{L}\right\}$, where $d\left ( a_{i},a_{j} \right )=\parallel \mathbf{p}_{i} - \mathbf{p}_{j} \parallel_{2}$, $\mathbf{p}_{i}$ and $\mathbf{p}_{j}$ denote the coordinates of $a_{i}$ and $a_{j}$, and $\theta$ is a hyperparameter representing distance threshold.
Then, we define the protein–ligand complex as an undirected graph $\mathcal{G}\left ( \mathcal{V},\mathcal{E} \right )$, where $\mathcal{V}=\mathcal{V}^{L} \cup  \mathcal{V}^{P}$ is the node set and $\mathcal{E}$ is the edge set containing two types of edges formed by atoms in $\mathcal{V}$, protein–ligand interactions and covalent bonds between ligand atoms. 

We introduce a distance threshold $\theta$  that can significantly reduce the size of the graph. Furthermore, we do not employ covalent bonds within the protein, which also greatly reduces the number of edges in the complex graph.
Then, we number the ligand atoms and retained protein atoms. According to the number, we construct the corresponding adjacency matrix $\mathbf{A}=\left [ A_{ij} \right ]_{\mathcal{N}^\mathcal{V}\times \mathcal{N}^\mathcal{V}}$, where $A_{ij}$ is defined as follows:
\begin{equation}
A_{ij}=\left\{\begin{matrix}
1, &   a_{i}, a_{j}\in \mathcal{V}^{L} \ \mathrm{and} \ \ a_{i}, a_{j} \ \mathrm{are \ connected} \ \mathrm{and} \ i\neq j \\
1, & a_{i}\in \mathcal{V}^{L}, a_{j}\in \mathcal{V}^{P} \ \mathrm{and} \ d\left ( a_{i},a_{j} \right ) \leq \theta  \\
0, & \mathrm{otherwise} \end{matrix}\right.
\end{equation}
By introducing the adjacency matrix representing both bond interactions and atomic nonbonded interactions, our model can learn how protein–ligand interactions affect the node features of each atom.

In the graph, each node includes 11 features, each feature is represented by a vector, and these vectors are concatenated to form the initial feature vector of a node. The types of edges include the covalent bonds between ligand atoms and protein–ligand interactions. The features of covalent bonds include the covalent bond type, whether the bond is in a ring, bond length, bond direction, and bond stereochemistry. To ensure that the dimensions of the two types of edges are consistent, the features of protein–ligand interactions are the same as those of covalent bonds, the bond length is the distance between two atoms, and other features take default values. In addition, two different types of edges are embedded in features using 0-1 codes to distinguish them. All features of an edge are encoded as vectors and concatenated to form the initial feature vector of an edge. A list of initial features for nodes and edges is summarized in Table 1.
\begin{table}
\footnotesize
  \caption{List of the initial features of nodes and edges.}
  \begin{tabular}{lll}
    \hline
    Type &	Name  &	Description
  \\
    \hline
   node features & atom type &	B, C, N, O, S, P, Se, Halogens, Metals, Other\\
    & atom charge number &	Formal charge for an atom. Range:[-5,5], Other\\
    & hybridization &	S, SP, SP2, SP3, SP3D, SP3D2, Other\\
 & atom valence	& Range:[0,7], Other\\
& atom degree &	Total number of bonded atom neighbors \\
 & & Range:[0,10], Other \\
& number of hydrogens &	Explicit and implicit hydrogens. Range:[0,8], Other\\
& atom coordinates & Position coordinates of atoms in 3D space \\
& chirality &  Unspecified, 
Tetrahedral$\_$CW, Tetrahedral$\_$CCW, Other	\\
& atomic mass &	Mass of a single atom\\
& aromatic & Whether if the atom is aromatic. 0 or 1 \\
& belongs to the protein & Whether the atom belongs to the protein, 0 or 1\\
\hline
edge features & covalent bond type &
Single, Double, Triple, Aromatic, Unspecified, Zero, Other \\
& aromatic & 	Whether the bond is in an aromatic ring. 0 or 1\\
& bond length &	Distance between connected atoms in 3D space \\
& bond direction & 
None, Endupright, Enddownright, Eitherdouble, Unknown	\\
& bond stereochemistry	 &  
Stereonone, Stereoany, Stereoz, Stereoe, Stereocis, Stereotrans  \\
& edge type & Protein–ligand interaction or a bond between ligand atoms.\\
 & & 0 or 1 \\
    \hline
  \end{tabular}
\end{table}

It is worth noting that the bond length in edge features is the Euclidean distance calculated based on the coordinates of $a_{i}$ and $a_{j}$ in 3D space, which is a continuous real value denoted by $d\left ( a_{i},a_{j} \right )$. 
To further simplify the computation and improve the model performance, we discretize the distance as shown in Eq. 2.
\begin{equation}
\hat{d}\left ( a_{i},a_{j} \right )=\left \lfloor d\left ( a_{i},a_{j} \right )\right \rfloor
\end{equation}
where $\hat{d}\left ( a_{i},a_{j} \right )$ denotes the value after discretization. SS-GNN applies the single undirected graph representation method based on the distance threshold, which greatly reduces the amount of computation and makes the model more lightweight.

\subsection{2.2 Hybrid Feature Extraction based on GNN-MLP}
Different from other methods, we propose a hybrid feature extraction module named GNN-MLP to extract the features of complexes. These two modules are independent of each other; the GNN-based network is applied to learn the latent features of atoms, and a multilayer perceptron (MLP) is applied to learn the latent features of edges. Each feature extraction module is very simple and lightweight.

\textbf{(1)	Node feature extraction based on GIN}

Xu et al. developed a simple and powerful graph learning method, the graph isomorphism network (GIN), and theoretically proved that the model has the maximum discriminant ability in GNNs\cite{xu2018powerful}. We utilize a GIN-based module to learn the node representations of the protein–ligand complex. The inputs of GIN are $\mathbf{x}$, $\mathbf{A}$, where $\mathbf{x}$ is the node initial feature vector and $\mathbf{A}$ is the corresponding adjacency matrix. First, the initial feature vectors of nodes are obtained; then, the representations of nodes are updated by aggregating information from neighbor nodes based on the adjacency matrix; and finally, iterative message passing is employed to extract the latent representations of the nodes. Since the composition of a function can be represented by an MLP, the MLP method is applied to update the node features in the GIN. GIN updates node features as follows:

\begin{equation}x_{i}^{'}= \mathrm{MLP}\left ( \left ( 1+\varepsilon \right )x_{i} +\sum _{j\in \mathcal{N}\left ( i \right )}x_{j}\right )\end{equation}
where $\varepsilon$  is either a learnable parameter or a fixed scalar.
We utilize an undirected graph to represent the protein–ligand complex, so the information for the protein atoms can also be updated from the ligand information.  

The GIN consists of two GIN layers, each of which is followed by a batch normalization layer to speed up the training. Node features and the adjacency matrix are fed into the GIN module to extract the latent features of the atoms of the complex, and the output $\mathbf{x}^{'}=\mathrm{GIN}\left (\mathbf{x},\mathbf{A} \right)$ is the latent feature vector. Only a single graph of protein–ligand complexes is fed into the GIN network, resulting in less input data, thereby reducing model computation. Only two GIN layers are applied in this module, resulting in a relatively lightweight model with only 0.039M parameters. 

\textbf{(2) Edge feature extraction based on MLP}

The MLP-based module is a multilayer feedforward neural network that consists of three layers, where the first two layers are followed by a ReLU activation function for nonlinear transformation. MLP1 is our edge feature extraction module (Figure 2(a)) designed to learn the edge features of the protein–ligand complex, which include two types: covalent bonds inside the ligand and edges connecting protein and ligand atoms.
The input $\mathbf{e}$ is the edge initial feature vector, and the output $\textbf{e}^{'}$ is the edge latent features: $\textbf{e}^{'}=\mathrm{MLP1}\left ( \textbf{e} \right )$. 

\subsection{2.3 Feature Aggregation and Affinity Prediction} 

\textbf{(1) Edge-based atom-pair feature aggregation}

To well represent the interactions in the complex, we propose an edge-based atom-pair feature aggregation module.
We obtain aggregated feature embeddings by aggregating each edge and the pair of atoms it connects. The specific method is to concatenate the latent features of atom pairs obtained by GIN with the latent features of edges between them obtained by MLP1, that is, to perform a simple concatenation of the three representation vectors:
\begin{equation}
\mathbf{x}_{ij} =\mathbf{x}_{i}^{'}\left| \right|\mathbf{x}_{j}^{'}
\end{equation}
\begin{equation}
\mathbf{AGG}_{ij} =\mathbf{e}_{ij}^{'}\left| \right|\mathbf{x}_{ij},  \  \forall \ e_{ij}\in \mathcal{E} 
\end{equation}
where $\mathbf{x}_{i}^{'}$ and $\mathbf{x}_{j}^{'}$ denote the latent feature vectors of two atoms connected by edge ${e}_{ij}$ obtained through GIN, || is a concatenation of two vectors, $\mathbf{e}_{ij}^{'}$ denotes the latent representation vector of the edge obtained by MLP1, and $\mathbf{AGG}_{ij}$ can be interpreted as the final information of aggregated features and is directly delivered to the affinity prediction module.

\textbf{(2)	Graph pooling-based affinity prediction}

In the last part, we utilize the MLP2 module and the graph pooling module for affinity prediction. The MLP2 module is a 4-layer feedforward neural network; except for the last layer, each layer is followed by a ReLU activation function for nonlinear transformation, and the fourth layer is the output layer. The aggregated feature embeddings representing the final states of edge-based atom pairs are passed through MLP2 to obtain the output value of each edge, which is finally used for affinity prediction, as shown in Figure 2(c). 
 The number of input edge-based aggregated feature embeddings is $\mathcal{N^{E}}$, where $\mathcal{N}^{\mathcal{E}}=\left|\mathcal{E} \right|$. The predicted value $\mathbf{y}$ is obtained through MLP2, which is an $\mathcal{N^{E}}$-dimensional vector, and each element in the vector represents the output for each edge. 
Finally, a graph pooling module is used to sum the output values of each edge as the binding affinity prediction of the complex.
\begin{equation}
y\left ( e_{ij} \right )=\mathrm{MLP2}(\mathbf{\mathbf{AGG}}_{ij}), \  \forall \ e_{ij}\in \mathcal{E}
\end{equation}
\begin{equation}
\hat{y}=\mathrm{ADDPOOL}\left ( \mathbf{y} \right )= \sum_{e_{ij}\in \mathcal{E}}y\left ( e_{ij} \right )
\end{equation}
where $\mathbf{AGG}_{ij}$ denotes the aggregated embedding of atom pairs based on edge $ e_{ij}$, $y\left ( e_{ij} \right )$ denotes the predicted output of an edge, $\mathcal{E}$ is the set of edges, ADDPOOL is the sum of all elements in the vector $\mathbf{y}$, and $\hat{y}$ is the final output value representing the binding affinity of the complex. 
Each module in the SS-GNN adopts a concise network structure, and the size of each module is shown in Table 2. A simplified single graph representation method based on a distance threshold and a simple feature extraction process leads to a lightweight model.
\begin{table}
\small
  \setlength\tabcolsep{14pt}
  \caption{The size of each module in the SS-GNN.}
  \begin{tabular}{llll}
    \hline
    Network module	& GIN  & MLP1 & MLP2
  \\
    \hline
network layers & 2 & 3 & 4 \\
parameters & 0.039 M & 0.003 M & 0.526 M  \\
    \hline
  \end{tabular}
\end{table}

\subsection{2.4 Loss Function}
In this end-to-end model SS-GNN, we treat the affinity prediction as a regression task. Given a dataset with $N$ samples, the predicted value and the ground truth of a certain sample are $\hat{y}_{i}$ and $y_{i}$, respectively. The loss function uses the MSE loss, defined by Eq. 8, and the training objective is to minimize the loss function. 
\begin{equation}
\mathrm{MSE \ loss}=\frac{1}{N}\sum_{i=1}^{N}\left ( \hat{y}_{i}-y_{i} \right )^{2}
\end{equation}

\section{3 Experiments}
In this section, we conduct a comprehensive experimental evaluation of the recent PDBbind v2016 and v2013 core sets to explain the benefits of exploiting the proposed model in affinity prediction. In the following subsections, we first introduce the distance threshold selection, analyze our proposed model through extensive ablation studies and then report experimental comparisons with recently proposed state-of-the-art methods. Finally, we present a detailed discussion of our experiments and provide useful insights and conclusions.

\subsection{3.1. Datasets and Evaluation Protocols }
To evaluate the performance of our proposed method, we adopt the widely used benchmark PDBbind dataset v2019\cite{wang2004pdbbind,wang2005pdbbind}. This dataset is a well-known public dataset used to predict  DTBA, and is a comprehensive database composed of 3D structure data of drug targets. This dataset provides 3D structures of protein–ligand complexes and the corresponding binding affinity represented by pKa values determined experimentally. It includes three overlapping subsets, namely, the general set $U_{g}$, the refined set $U_{r}$ and the core set $U_{c}$, where $U_{c}\subset U_{r}\subset U_{g}$. The general set contains all samples of the dataset, while the refined set is a subset with higher quality data selected from the general set. The core set is designed as the highest quality benchmark and is often used as a test set. The protein–ligand complexes in the core set have high-quality crystal structures and reliable experimental affinity data. 

In this paper, we employ two test sets (the v2016 and v2013 core sets) to test the performance of SS-GNN. The v2016 core set\cite{su2018comparative} contains 285 structurally diverse ligand–receptor complexes (270 samples are used for testing, and 15 samples fail in the reading and processing of protein or compound structure information). The v2013 core set\cite{li2014comparative} contains 195 complex samples (189 samples are used for testing, and 6 samples fail in the reading and processing of protein or compound structure information).

For the experiments with the v2016 core set, we remove the overlapping part of the corresponding core set from the refined set, and the remaining samples are employed for model learning, of which 90$\%$ are used as the training set (4,073 samples) and 10$\%$ are used as the validation set. Moreover, we carry out a supplementary experiment on the larger general set to further analyze the generalizability of our model. Samples in the general set that overlap with the core set are removed. Similar to the process with the refined set, 90$\%$ of the remaining samples are used as the training set (15,394 samples), and 10$\%$ are used as the validation set. The processing of the experimental dataset for the v2013 core set is the same as that of the v2016 core set. In the end, 4,005 training samples are obtained in the refined set, and 15,317 training samples are obtained in the general set.

For model evaluation, we follow previous work to evaluate performance from different perspectives using two main indicators: root mean squared error (RMSE)\cite{wackerly2014mathematical} and Pearson correlation coefficient ($R_{p}$). In addition, to achieve a more diverse evaluation, the concordance index(CI)\cite{gonen2005concordance}, coefficient of determination ($R^{2}$) and mean absolute error (MAE) are calculated.
The smaller the values of RMSE and MAE and the larger the values of $R_{p}$, $R^{2}$ and CI are, the better the performance of the model.

\subsection{3.2 Implementation Details}
We implement our approach based on the PyTorch toolbox. Experimentally, we apply the Adam optimizer and set the learning rate to 0.001. We train our network for 1000 epochs, and in each epoch, the training set is randomly divided into 192 mini-batches. Modeling experiments and benchmarking are carried out on a machine with an Intel(R) Xeon(R) CPU E5-2678 v3 @ 2.50 GHz CPU and an NVIDIA GeForce RTX 2080Ti graphics card.

\subsection{3.3 Distance Threshold Selection based on Cross-validation}

In this subsection, we introduce the method of distance threshold selection and verify the necessity of distance threshold selection.
We choose the refined set as the training set and perform 5-fold cross-validation separately for different thresholds ranging from 4 Å to 8 Å.
Table 3 shows the $R_{p}$ values of the model under different thresholds on the PDBbind v2019 refined set.
The experimental results show that the model performs the worst when the threshold is 4 Å, while the difference in $R_{p}$ is not significant when the threshold is 5 Å and larger.
\begin{table}
  \caption{Cross-validation experimental results with different thresholds.}
  \begin{tabular}{llllll}
    \hline
   Threshold &	4 Å & 5 Å &	6 Å	 & 7 Å &	8 Å \\
$R_{p}$ (↑) &	0.674±0.033	& 0.717±0.021 & 	0.716±0.026 & 0.711±0.022 &	0.710±0.026 \\
    \hline
  \end{tabular}
\end{table}

To verify the effect of threshold selection on the size of the constructed complex graph, we calculate the average values of the number of atoms and edges for the 285 complex samples in the v2016 core set at different thresholds, as shown in Figure 3.
\begin{figure}
   \centering
   \includegraphics[width=16cm]{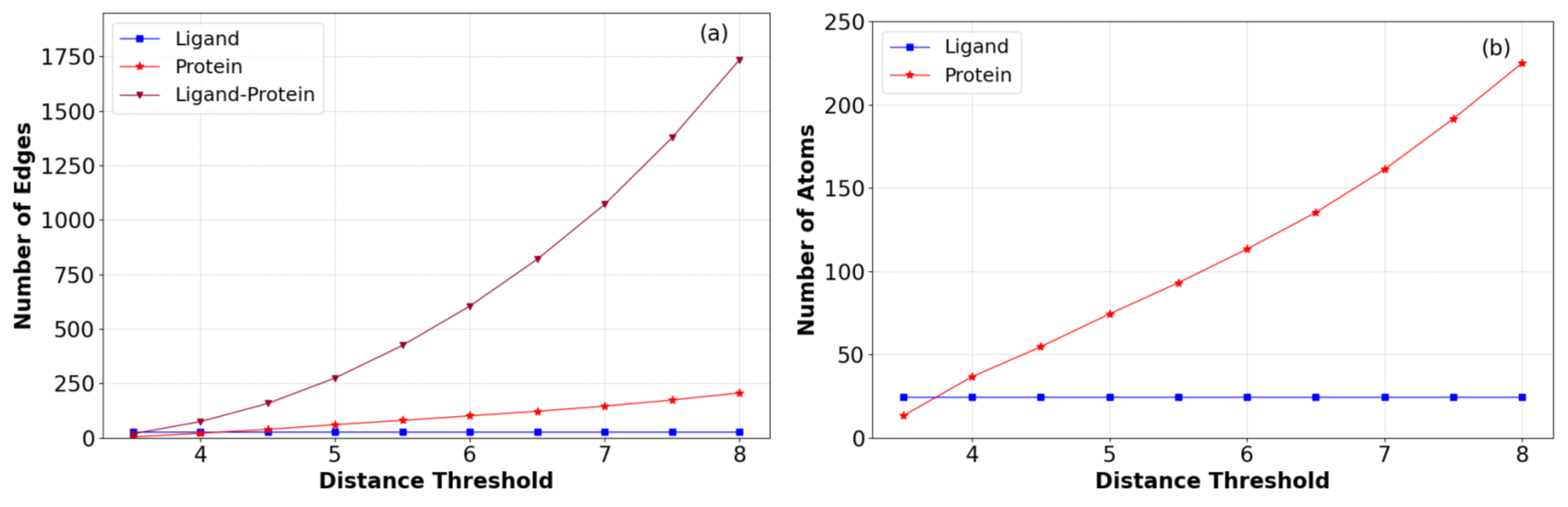}
  \caption{\textbf{Average values of the number of edges (a) and atoms (b) for the 285 samples in the v2016 core set at different thresholds.}}
\end{figure}
As in common practice, all water molecules and hydrogen atoms in the PDB structures are removed. The number of ligand atoms and the intramolecular edges do not change with increasing distance threshold. With an increase in threshold, the number of protein atoms increases linearly, and the number of edges between ligands and proteins also rises dramatically. Figure 4 shows the number of protein covalent bonds for 50 samples randomly selected from the 285 samples in the PDBbind v2016 core set. 
\begin{figure}
   \centering
   \includegraphics[width=16cm]{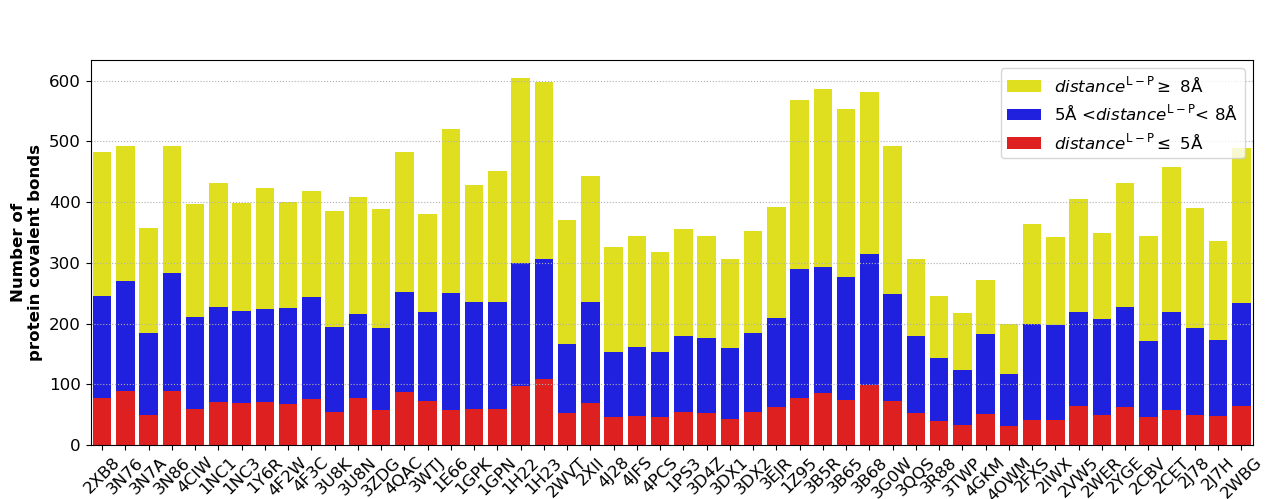}
  \caption{\textbf{Number of covalent bonds formed by atoms that satisfy the threshold condition in the protein, where $\mathbf{distance^{\mathrm{L-P}}}$ is the distance between ligand atoms and protein atoms.}}
\end{figure}
The number of protein covalent bonds increases significantly with increasing distance threshold. However, our model does not use any covalent bonds between proteins, which is markedly different from other models. Figure 5 depicts the number of ligand–protein connections for 50 randomly selected samples. The average number of connections at a distance threshold of 5 Å is reduced to 1/6 of that at 8 Å.  
\begin{figure}
   \centering
   \includegraphics[width=16cm]{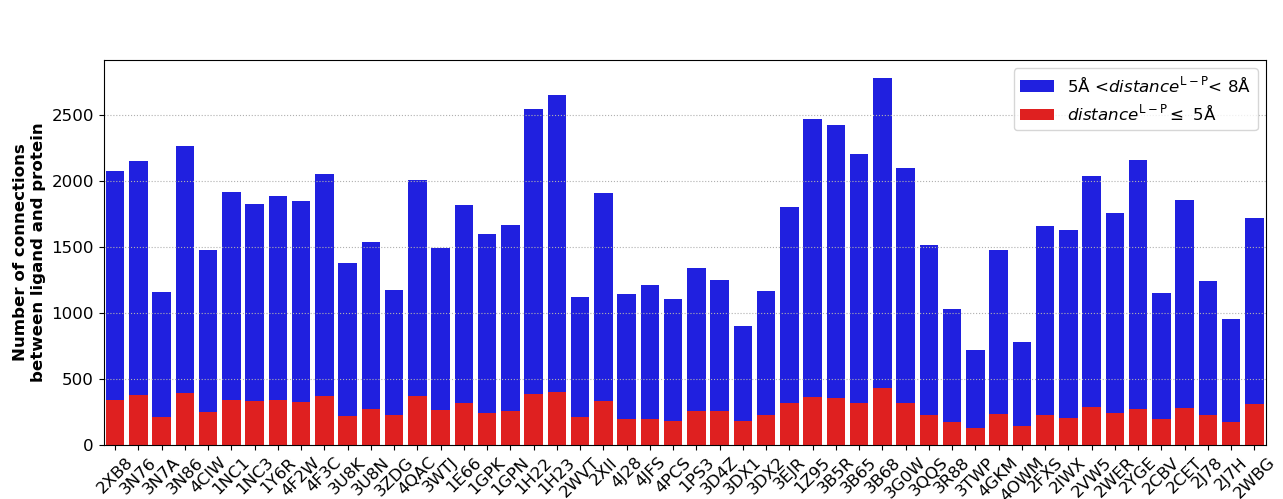}
  \caption{\textbf{Number of connections between ligand atoms and protein atoms satisfying a threshold condition, where $\mathbf{distance^{\mathrm{L-P}}}$ is the distance between ligand atoms and protein atoms.}}
\end{figure}
Considering the prediction performance and computational cost, we finally select 5 Å as the distance threshold. The subsequent experiments are carried out under the 5 Å threshold. Compared with other methods, SS-GNN applies a complex graph representation method based on a distance threshold, resulting in a substantial reduction in the size of the graph, which fully illustrates the computational advantages of SS-GNN.

To further illustrate the efficiency of SS-GNN, we test the model forward propagation runtime on the PDBbind v2019 refined set. When the threshold is 8 Å, the average prediction time per sample is 0.7 ms, and when the threshold is 5 Å, the average prediction time per sample is 0.2 ms. The lightweight model architecture and concise data processing procedure result in an efficient model.

\subsection{3.4 Ablation Studies}
To better understand the contribution of each component in the model to the overall performance, we remove each component from the model and conduct the ablation experiments using the PDBbind v2016 refined set, and the results are shown in Table 4. 
\begin{table}
  \footnotesize
  \caption{ Experimental results showing the effect of different components on the model. In each table cell, the mean value over five runs is reported as well as the standard deviation.}
  \begin{tabular}{llllll}
    \hline
    Architecture &	RMSE (↓) &	$R_{p}$ (↑) &	CI (↑) &	$R^{2}$ (↑) &  MAE (↓)  \\
    \hline
  SS-GNN	& \textbf{1.181±0.047}	& \textbf{0.853±0.012} & \textbf{0.833±0.006} &\textbf{	0.701±0.024} & \textbf{0.920±0.035}
  \\
   SS-GNN$_{\mathrm{remove \ MLP1}}$  &	1.184±0.035 & 0.845±0.011 & 0.827±0.005 & 0.700±0.018 &	0.927±0.028
 \\
SS-GNN$_{\mathrm {remove \ DDA}}$ &	1.194±0.052&	0.849±0.006&	0.828±0.005 & 0.694±0.027 &	0.926±0.048
\\
SS-GNN$_{\mathrm {1-layer \ GIN}}$ &	1.185±0.046&	0.849±0.011	&0.829±0.005 & 0.699±0.024 &	0.928±0.045
\\
SS-GNN$_{\mathrm{3-layers \ GIN}}$ &	1.220±0.018&	0.838±0.007&	0.825±0.005 & 0.682±0.01 &	0.942±0.026
\\
    \hline
  \end{tabular}
\end{table}

First, we evaluate the usage of MLP1 in the GNN-MLP module, which is a fully connected neural network for learning the latent features of edges. Compared with SS-GNN$_{\mathrm{remove \ MLP1}}$ (with MLP1 module removed), SS-GNN has a 0.9$\%$ increase in $R_{p}$, a 0.3$\%$ decrease in RMSE, and a 0.7$\%$ increase in CI.
This shows that the MLP1 module can learn the latent representation of the interactions between atom pairs at a deeper level and improve the model performance to a certain extent.

Second, we evaluate the effect of discretization of the distances between atoms (DDA) on the model performance.
Compared with the model SS-GNN$_{\mathrm {remove \ DDA}}$ without distance discretization, the RMSE of SS-GNN is reduced by 1.1$\%$, $R_{p}$ is improved by 0.5$\%$, and CI is improved by 0.6$\%$. 
The results show that the DDA module is necessary for SS-GNN.

Finally, we evaluate the effect of the number of layers of GIN. SS-GNN using a 2-layer GIN shows an advantage over the model SS-GNN$_{\mathrm {1-layer \ GIN}}$ using 1-layer GIN (RMSE is reduced by 0.3$\%$, $R_{p}$ is improved by 0.5$\%$ and CI is improved by 0.5$\%$); however, the improvement is minor. Moreover, SS-GNN outperforms SS-GNN$_{\mathrm{3-layer \ GIN}}$ using a 3-layer GIN (RMSE is reduced by 3.2$\%$, $R_{p}$ is improved by 1.8$\%$, and CI is improved by 1.0$\%$), indicating that increasing the number of GIN layers does not always lead to better performance of the model.

\subsection{3.5 Comparison with the State of the art}
In this subsection, we first test our model on the PDBbind v2016 core set and then compare the proposed approach with other state-of-the-art methods on two datasets. 

\textbf{Experiments on the PDBbind core set. }We employ the general set and refined set in PDBbind for model training and test the model on the PDBbind v2016 core set. The results are shown in Table 5. 
\begin{table}
\scriptsize
  \caption{Results of PDBbind dataset experiments.}
  \begin{tabular}{llllllll}
    \hline
    Type & Test set & Training set & RMSE (↓) & $R_{p}$ (↑) &	CI (↑)  & $R^{2}$ (↑) & MAE (↓) \\
    \hline
  SS-GNN$_\mathrm{best}$ &v2016/270  & 4073  & 1.289 & 0.832 & 0.819 & 0.645 &	1.011
\\
    &  & 15394 & 1.128 &	 0.870 & 0.839 & 0.728 &	0.902\\
      & v2013/189  & 4005 &
      1.355 &	0.803 &	0.803 &	0.634 &	1.094
 \\
    &   & 15317 & 1.296  &	0.831  &	0.816 &	0.665 	& 1.026 

 \\
    SS-GNN$_\mathrm{average}$ &v2016/270  & 4073 & 1.281±0.021 & 0.822±0.006 & 0.813±0.004 & 0.649±0.011 &	1.012±0.016
\\
      & & 15394 & 1.181±0.047 & 0.853±0.012	 &	0.833±0.006 & 0.701±0.024 &	0.920±0.035
\\
     & v2013/189   & 4005 & 1.454±0.05 &	0.795±0.008 &	0.798±0.010 &	0.578±0.029 & 1.165±0.055
 \\
  &  & 15317 & 1.347±0.049 & 0.816±0.012 &	0.808±0.007 & 0.638±0.027 &	1.074±0.031
 \\
    \hline
  \end{tabular}
\end{table}
All experiments in this paper are repeated 5 times with different random seeds. Each random seed represents a random shuffle of the dataset.
In each experiment, 90$\%$ of the samples are randomly selected as the training set, and the remaining 10$\%$ are selected as the validation set for model selection. 
We finally take the mean and standard deviation of the results of five independent experiments as the result of the average model SS-GNN$_\mathrm{average}$ and take the model result with the largest  $R_{p}$ value as the result of the best model SS-GNN$_\mathrm{best}$.
For the PDBbind v2016 core set, the $R_{p}$ of the best model trained on the refined set reaches 0.832, and that of the average model is 0.822; for the general set, the $R_{p}$ of the best model reaches 0.870, and that of the average model is 0.853. The model achieves good performance on the refined set with a small sample size. Nonetheless, with the expansion of the training dataset, the performance of the model is greatly improved, which further expands the prediction advantage. For the PDBbind v2013 core set, the $R_{p}$ of the average model trained on the general set reaches 0.816.

To better represent the findings, the predicted binding affinities obtained using the PDBbind v2016 core set are shown in Figure 6, which presents the test results for the best models trained on the general and refined sets based on the PDBbind v2016 core set. The predicted values are highly correlated with the ground truth values.
To ensure the stability of model prediction performance, 5 different random seeds are used in the model experiments in this paper.

\begin{figure}
   \centering
   \includegraphics[width=16cm]{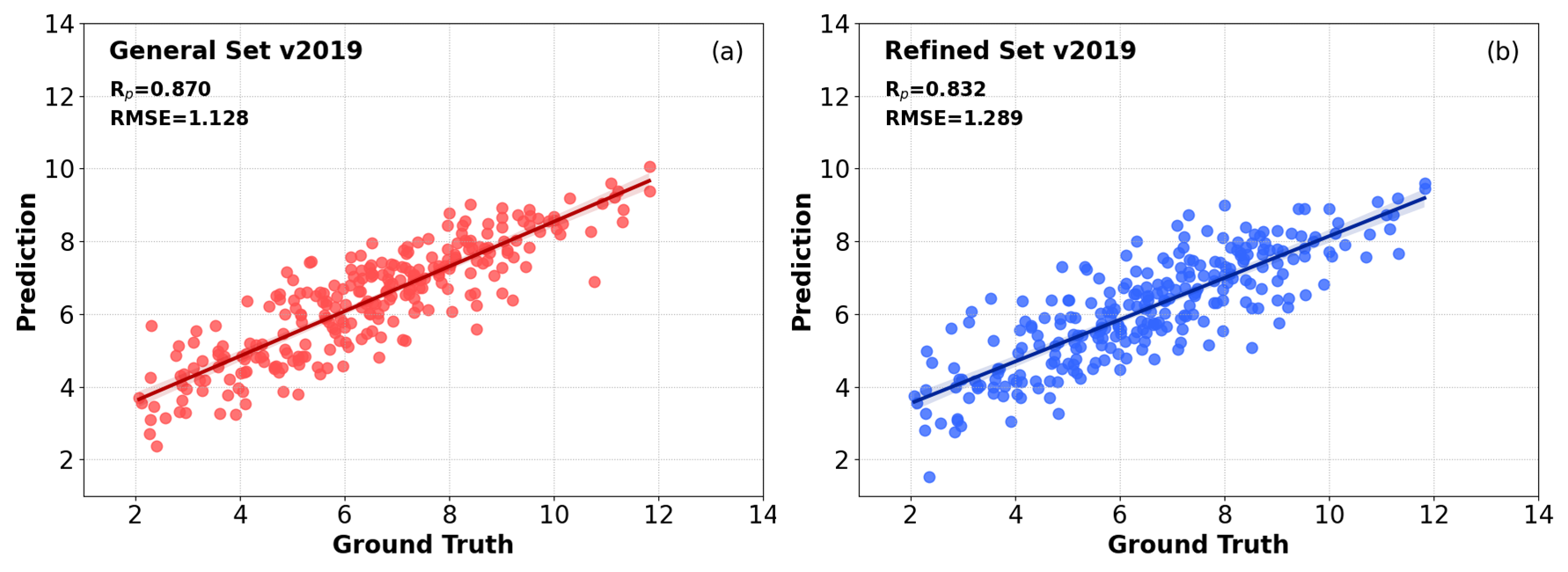}
  \caption{\textbf{Correlation plot for the PDBbind v2016 core set given by the best SS-GNN models trained on (a) the general set and (b) the refined set.}}
\end{figure}

\textbf{Comparison with 9 state-of-the-art methods.} We compare our proposed method with state-of-the-art methods, including Pafnucy\cite{stepniewska2018development}, $\mathrm{K_{Deep}}$\cite{jimenez2018k}, OnionNet\cite{zheng2019onionnet}, IGN\cite{jiang2021interactiongraphnet}, SIGN\cite{li2021structure},  HPC/HWPC\cite{liu2021hypergraph}, AGL-Score\cite{nguyen2019agl}, 
$\Delta \textrm{VinaRF}_{20}$\cite{wang2017improving,su2018comparative} and FAST\cite{jones2021improved}. Among them, Pafnucy, $\mathrm{K_{Deep}}$, and OnionNet are CNN-based models, and IGN and SIGN are GNN-based models. FAST is a model fusion of GNN and CNN. AGL-Score, HPC/HWPC, and $\Delta \textrm{VinaRF}_{20}$ are ML-based models.
Table 6 compares the results of our proposed SS-GNN with those of the state-of-the-art methods for the PDBbind core set v2016.
SS-GNN ranks first on the general set, with a 2.4$\%$ improvement in $R_{p}$ over the current best result, and presents a 5.2$\%$ improvement compared to the current most advanced GNN-based approach. When learning with very limited samples, SS-GNN outperforms similar DL-based methods, which demonstrates that our lightweight structure can effectively learn deep features of the interactions of protein–ligand complexes. We select the chemical and biological attributes that can well represent the atom information during the data processing procedure and introduce an edge-based atom-pair feature aggregation module, which can better represent the interactions between atoms. We further utilize a GIN-based network and an MLP to learn the latent features of nodes and edges, respectively, in the complex graph. Therefore, despite the low number of atoms and interactions employed by SS-GNN, the model still achieves good performance. As the amount of training data increases, our model can provide more accurate predictions.
\begin{table}
\scriptsize
  \caption{Performance comparison using the PDBbind v2016 core set and v2013 core set.}
  \begin{tabular}{llllllll}
    \hline
     & &\multicolumn{3}{c}{PDBbind v2016 core set }&\multicolumn{3}{c}{PDBbind v2013 core set } \\
    Architecture & Training samples & Test samples &  RMSE (↓) & 	$R_{p}$ (↑) & Test samples & RMSE (↓) & 	$R_{p}$ (↑)\\
    \hline
    Pafnucy & 11906 &	290	& 1.420 &	0.780 &	195 &	1.620 &	0.700
 \\ 
    SIGN &3767&	290	&1.316&	0.797&	-&	-&	-
\\
FAST&11717&	290&	1.308&	0.810&	-&	-&	-
\\
IGN	& 8298 & 262 &1.291\textsuperscript{\emph{b}}&	0.811\textsuperscript{\emph{b}} & - & - & -
\\
$\Delta \textrm{VinaRF}_{20}$ &	3336 &	285&	- &	0.816 &	195 &	- &	0.686
\\
OnionNet &	11906 &	290	& 1.278 &	0.816 &	108	& 1.503	& 0.782
\\
$\mathrm{K_{Deep}}$ & 3767×24\textsuperscript{\emph{a}} &	290 & 1.270 &	0.820 &	- &	- &	-
\\
HPC/HWPC &	3772 &	285	& 1.307	& 0.831 & 195/2764	& 1.483 &	0.784
\\
AGL-Score & 3772 &	285	& 1.271	& 0.833	&195/3516 &	- &	0.792
 \\
$\mathrm{SS-GNN_{refined\ set}}$ &	4073 &	270	& 1.249 &	0.821 &	189/4005 &	1.454  &	0.795
\\
$\mathrm{SS-GNN_{general\ set}}$  &15394 &	270 &\textbf{1.181} &	\textbf{0.853} &	189/15317 &	\textbf{1.347}&	\textbf{0.816}
\\
    \hline
  \end{tabular}

  \textsuperscript{\emph{a}} The datasets of $\mathrm{K_{Deep}}$ were augmented 24 times by rotation;
  \textsuperscript{\emph{b}} The results of IGN are the indicators of the average model.
\end{table}

AGL-Score (based on algebraic graph descriptors) and HPC/HWPC (based on a hypergraph topology framework) achieve better results with less data ($R_{p}$=0.833 and $R_{p}$=0.831). 
ML-based methods rely more on expert knowledge and can achieve excellent results under reasonable feature extraction. As the amount of data increases, DL methods have greater potential.
We also test the efficiency of HPC/HWPC on the PDBbind v2019 refined set. The average prediction time per sample is 1.2$\times 10^{4}$ ms (implemented with our optimized code which is orders of magnitude faster than the original implementation), while SS-GNN only needs 0.2 ms. The feature extraction process of HPC/HWPC is complicated, computationally intensive and slow. 
Due to the lack of large-scale standard datasets, our proposed model has not been tested on very large-scale datasets, but its superiority in accuracy and efficiency makes it more suitable for large-scale molecular docking tasks.

\section{4 Conclusion}
In this paper, we have proposed a novel simple-structured graph neural network model (SS-GNN) for drug-target binding affinity (DTBA) prediction. We utilize the single undirected graph representation method based on the distance threshold to reduce the size of the complex molecular graph, thereby reducing the computational complexity of the model. The process of feature extraction and affinity prediction is straightforward. The concise graph representation and simple model architecture improve the efficiency of SS-GNN. Experiments confirm the superiority of SS-GNN, which significantly outperforms state-of-the-art methods on the PDBbind dataset. However, it has not been verified which of the chemical properties we input are critical in constructing the complex graph. In addition, whether the covalent interactions between protein atoms have an effect on the interactions of the complex needs further verification.

\begin{acknowledgement}
This research was partially supported by the National Social Science Fund of China (No. 18ZDA200), the Hebei Provincial Key Research and Development Project of China (No. 20370301D), and the Key Technology Development Project of Hebei Normal University (No. L2020K01).
\end{acknowledgement}





\bibliography{achemso-demo}

\end{document}